# A LATENT SURVIVAL ANALYSIS INTEGRATED SIMULATION PLATFORM FOR NURSING HOME STAFFING STRATEGY EVALUATION

Xuxue Sun[1], Nan Kong[2], Nazmus Sakib[1], Chao Meng[3], Kathryn Hyer[4], Hongdao Meng[4], Chris Masterson[5], Mingyang Li[1]*

[1]Department of Industrial and Management Systems Engineering
University of South Florida, Tampa, FL 33620, USA
[2]Weldon School of Biomedical Engineering
Purdue University, West Lafayette, IN 47907, USA
[3]School of Marketing, University of Southern Mississippi
[4]School of Aging Studies
University of South Florida, Tampa, FL 33620, USA
[5]Greystone Health Network, Tampa, FL 33610, USA

## 1. Introduction

Skilled nursing facilities, or nursing homes (NHs), are responsible for caring the frail and elderly population, who may suffer from multiple chronic diseases, multi-functional (e.g., physical, mental and social) losses and aging-related disabilities, by providing 24/7 personal medical care and daily-living assistance. Due to rapid population aging, the United States (U.S.) will soon experience a significant growth in the elderly population with disability and the trend will continue in future decades.

While NH care is known to be expensive, the large spending does not necessarily translate into high quality of care at NHs. Over 120,000 deficiencies were issued to NHs due to regulatory violations in 2013 [1], and 20.5% of NHs received deficiencies issued for causing potential/actual harm to their residents in 2014 [2]. Moreover, rapid population aging triggers escalating workforce shortage and public financing shortfall [3]. The above challenges imply that providing high-quality NH care is clearly at risk.

NH staffing is influential to resident outcomes (e.g., re-hospitalization, morbidity and mortality rates, percentages of pressure sores and incontinence) and service quality indices of NH facility (e.g., facility deficiency citation) [4, 5]. So far, practice at NH is either based on diverse administrator experience [6] or "one-size-fit-all" government regulation (e.g., minimum staff-to-resident ratio requirements enforced by federal/state agencies) [7]. There is a lack of predictive analytics integrated simulation-based decision support platform to inform appropriate staffing decisions to NH administrators in various practical contexts. On the other hand, in the existing literatures of simulation modeling for healthcare resource (e.g., bed capacity, staffing, etc.) planning and management, most of them focused on acute care settings, such as hospital [8-12] and emergency department [13-15]. There were also many operation research studies utilizing simulation to investigate resource management in other service systems, such as call center [16], transportation system [17] and power system [18].

To investigate appropriate NH staffing decisions, there are several challenges to overcome. First, unlike patients in acute-care hospitals where their length-of-stays (LOSs) do not differ much, NH residents either spend a few days to a month in the facility receiving post-acute care to recover from surgery or stay in the facility for a substantially longer time to maintain their functioning performance on daily activities (e.g., eating, dressing, bathing) [19]. Thus, their NH LOS, or dwelling duration, may vary from a few weeks to several months and even years. The highly skewed and right-censored LOS observations make conventional LOS modeling assumptions of symmetry and normality inappropriate in many simulation-based healthcare literature [20]. In addition, NH residents have multiple potential discharge dispositions. They may be either discharged to their residential community for further recovery or transferred to hospital due to occurrence of critical events (e.g., fall and infection). The mutually exclusive events are community discharge and re/hospitalization; and whichever comes first determines the NH dwelling duration of a resident. Existing healthcare literatures of simulation modeling have studied various parametric survival analysis models (e.g., Lognormal [8], Weibull [21], and exponential models [9]) to predict LOS, but they cannot address the data complexity resulting from multiple types of discharge dispositions. There is a need to develop predictive analytics method to characterize the complex LOS data via considering multiple discharge dispositions and to improve LOS prediction of NH residents. Further, there is a need to integrate the successful predictive model with computer simulation to improve resident flow modeling and prediction performance at NH facility level.

Moreover, NH residents often suffer from diverse chronic diseases and functional limitations, and their service needs are highly heterogeneous. The service demand heterogeneity of NH residents cannot be well addressed by existing approaches. Many of existing studies utilized computer simulation to model service demand of patients for healthcare resource planning and management, such as appointment scheduling [10, 14], surgery planning [13, 21], capacity planning and bed management [20, 22-23], workforce planning and staffing management [11-12, 24]. However, many of these works mainly assumed homogeneity among different individual patients and analyzed service demand based on patient volume, which neglected the diversity in the amount of care needed by each individual resident with varied individual characteristics (e.g., multiple chronic diseases, physical, mental, cognitive functioning limitations and losses, etc.) on different types of caregivers (e.g., nurses, aides, etc.). For example, according to Centers for Medicare & Medicaid Services (CMS) [25], average daily time spent by certified nursing assistants on caring residents varies from 30 to 240 minutes, depending on each resident's functional performance. With the same patient volume but different patient censuses, the total service demand at the facility level can differ drastically. There is a need to develop a simulation model to characterize the highly heterogeneous service demand of individual NH residents with improved simulation modeling fidelity.

To address the aforementioned research needs and to fill the research gaps, we propose a latent survival analysis integrated computer simulation platform to characterize the heterogeneous service demand of NH residents over time and to further allow staffing strategy evaluation under various census composition scenarios of NH residents. The main contributions of this paper lie in the following three aspects. First, as compared to the existing literatures of integrating survival analysis in simulation modeling for healthcare applications, the proposed predictive analytics method based on latent survival analysis can characterize multiple discharge dispositions (e.g., community discharge, hospitalization) of NH residents with improved NH LOS modeling performance. Further, the development of sampling algorithm for the proposed LOS model enables the integration of developed predictive analytics with computer simulation to estimate service demand of a population of NH residents at facility level. Second, as compared to the existing literatures of simulation-based service demand modeling in healthcare service operations and management, the proposed simulation model can characterize the heterogeneous service demand of NH individual resident via incorporating rich and diverse individual characteristics (e.g., functional performance, clinical diagnoses, care services received, treatment prescribed, etc.). The successful modeling of heterogeneous service demand can facilitate the staffing strategy evaluation for NH management group under various what-if scenarios of the census composition of a heterogeneous population of NH residents, and can further inform NH administrators about the most appropriate staffing strategy which succeeds in meeting the overall service demand over time at the most reduced total labor costs. Last but not least, as compared to the limited existing simulation-based healthcare applications in long-term care settings of NH research and real practice, the developed simulation-based staffing decision support platform is the first initiative in NH industry to integrate analytics-based method with computer simulation and NH domain knowledge in helping inform NH management group of NH staffing decision. In a proof-of-the-concept study, we used de-identified clinical assessment data from a local NH facility in Tampa Bay area, to illustrate the proposed work and to demonstrate its validity and superior performance. To enhance the tool usability, we design a graphic user interface so NH administrators can flexibly adjust practical contexts, modify staffing strategies, and visualize detailed simulation results, all of which help large-scale adoption of the tool in practice.

## 2. Methods

The proposed predictive analytics integrated simulation platform for NH staffing strategy evaluation was developed in a bottom-up manner. The proposed platform embraced several key modules, such as individual NH LOS predictor, individual daily staff-time simulator and graphical user interface. The details were elaborated as follows.

### 2.1 Individual NH LOS Predictor

To implement this module, historical data on NH LOS was used to develop and parameterize disposition-specific discharge probability distributions. These distributions were then used in the module to determine the LOS of each resident. With the predicted LOS for each resident, we determined whether the resident stays at NH on a given day after his/her arrival.

Each NH resident has multiple types of discharge disposition. S/he can either be discharged to home and community due to improved health condition or be transferred to a hospital due to occurrence of critical events (e.g., injury and infection) and/or deteriorating functioning status. In addition, multiple discharge events with different discharge dispositions are mutually exclusive and competing to each other. Whichever occurs first will determinate the dwelling duration of an individual NH resident and make the other events unobserved. For instance, a resident who is discharged to community still has a risk of being transferred to hospital. Since the hospitalization event has not been triggered when discharge-to-community event occurs first for a resident, time-to-hospitalization then becomes a latent variable that cannot be directly observed. Many existing literatures of survival analysis models in healthcare simulation either assumed single discharge disposition for LOS modeling [8-9, 21], or analyzed LOS data by segregating LOS observations into multiple groups with observed disposition labels [26]. Simply adopting them without considering latent discharge events and multiple discharge dispositions may lead to inaccurate LOS prediction of NH residents.

To address the above complexities of NH LOS data, we developed a predictive model based on latent survival analysis to account for multiple discharge dispositions and latent discharge events. Specifically, considering $N$ NH residents and each of them may be discharged to one of $K$ different disposition settings, let us denote random variable $T_{i\mu}$ to be the latent time-to-discharge of resident $i$ with disposition $\mu$, $i=1,...,N$, $\mu=1,...,K$. The NH LOS of resident $i$ can be expressed as $T_i^{min} = \min\{T_{i1}, ..., T_{iK}\}$. Then the disposition-specific discharge risk, $d_\mu(t)$, was modeled as

$$d_\mu(t) = \lim_{\Delta t \to 0} \frac{\Pr(t \leq T_{i\mu} \leq t+\Delta t \mid T_i^{min} \geq t)}{\Delta t}, \qquad (1)$$

where $d_\mu(t)$ captures the instantaneous probability of being discharged to disposition $\mu$ at time $t$. Thus, the overall instantaneous probability of being discharged was estimated by $d(t) = \sum_\mu d_\mu(t)$. Following Eq. 1, we expressed the cumulative probability of a resident staying in NH until time $t$, i.e., $S(t) = \Pr(T_i^{min} \geq t) = \exp[-\int_0^t \sum_\mu d_\mu(s)ds]$, and the cumulative probability of being discharged to disposition $\mu$ until time $t$, i.e., $F_\mu(t)=\Pr(T_\mu \leq t) = \int_0^t d_\mu(\tau)\exp[-\int_0^\tau \sum_\mu d_\mu(s)ds]d\tau$. We implemented the above mathematical expressions in the module to simulate how long each resident would stay in NH.

To account for right skewness in real LOS observations, we considered a log-normal distribution for each discharge disposition $\mu$ and calculated the disposition-specific discharge risk as $d_\mu(t) = \phi((\ln t - \eta_\mu)/\sigma_\mu) / [t\sigma_\mu \Phi(-(\ln t - \eta_\mu)/\sigma_\mu)]$, where $\eta_\mu$ and $\sigma_\mu$ represent discharge disposition-specific mean and standard deviation parameters, and $\phi(\cdot)$ and $\Phi(\cdot)$ represent the probability density function and cumulative distribution function of a standard normal distribution, respectively. We then fitted the LOS observations to estimate the above model parameters. We denoted binary indicators $Z_{i\mu}$'s, $\mu=1,...,K$, which takes value 1 if resident $i$ is discharged to disposition $\mu$, and 0 otherwise. To account for observations being right-censored, we further introduced a right-censored indicator $\delta_i$ for each LOS observation $t_i$, i.e., $\delta_i=1$ if $t_i$ is observed (i.e., resident $i$ is discharged) and $\delta_i=0$ if $t_i$ is right-censored (i.e., resident $i$ is still in NH by the end of the time period under investigation). We denoted **D** to be the set of observations, i.e.,

$\mathbf{D} = \{t_i, Z_{i\mu}, \delta_i, i = 1, \ldots, N, \mu = 1, \ldots, K\}$ and $\mathbf{\Theta}$ to be the set of unknown model parameters, i.e., $\mathbf{\Theta} = \{\mathbf{\Theta}_\mu\}_{\mu=1}^K$, where $\mathbf{\Theta}_\mu = \{\eta_\mu, \sigma_\mu\}$, then we computed the joint likelihood as

$$L(\mathbf{\Theta} \mid \mathbf{D}) = \prod_{i=1}^N \left\{ \prod_{\mu=1}^K \left( d_\mu(t_i) \exp[-\int_0^{t_i} \sum_\mu d_\mu(s) ds] \right)^{Z_{i\mu}} \right\}^{\delta_i} \left[ \exp[-\int_0^{t_i} \sum_\mu d_\mu(s) ds] \right]^{1-\delta_i}. \quad (2)$$

When all $\mathbf{\Theta}_\mu$'s are mutually exclusive or in other words, each disposition-specific risk model is characterized by a distinct set of model parameters associated with that discharge disposition, Eq. (2) can be further decomposed into the multiplication of multiple disposition-specific likelihoods, i.e., $L(\mathbf{\Theta} \mid \mathbf{D}) = \prod_{\mu=1}^K L_\mu(\mathbf{\Theta}_\mu \mid \mathbf{D})$, where $L_\mu(\mathbf{\Theta}_\mu \mid \mathbf{D})$ can be explicitly written as

$$L_\mu(\mathbf{\Theta}_\mu \mid \mathbf{D}) = \prod_{i \in I_\mu} \left\{ \frac{1}{t_i \sigma_\mu \sqrt{2\pi}} \exp\left[-\frac{(\ln t_i - \eta_\mu)^2}{2\sigma_\mu^2}\right] \right\}^{\delta_i} \prod_{i=1}^N \left[ 1 - \Phi\left(\frac{\ln t_i - \eta_\mu}{\sigma_\mu}\right) \right]^{1-\delta_i} \quad (3)$$

where $I_\mu$ is an index set of individuals who have been discharged to disposition $\mu$. To estimate the model parameters, we employed the maximum likelihood principle, e.g., $\widehat{\mathbf{\Theta}}_\mu = \text{argmin}_{\mathbf{\Theta}_\mu} - l_\mu(\mathbf{\Theta}_\mu \mid \mathbf{D})$, where $l_\mu(\mathbf{\Theta}_\mu \mid \mathbf{D})$ is the logarithm of Eq. (3) by ignoring the additive constant, i.e., $l_\mu(\mathbf{\Theta}_\mu \mid \mathbf{D}) = \sum_{i \in I_\mu} \left( -\ln \sigma_\mu - \frac{(\ln t_i - \eta_\mu)^2}{2\sigma_\mu^2} \right) - \sum_{i \notin \cup_\mu I_\mu} \ln\left(1 - \Phi\left(\frac{\ln t_i - \eta_\mu}{\sigma_\mu}\right)\right)$. The union set $\cup_\mu I_\mu$ is an index set of all individuals who have been discharged. We further employed the Newton-Raphson algorithm to realize the model estimation. The proposed algorithm for estimating the disposition-specific model parameters can be summarized as follows.

**Step 1**: At iteration $\tau=0$, initialize $\eta_\mu^{(0)}, \sigma_\mu^{(0)}, \mu = 1 \ldots K$, and further specify a small tolerance value $\varepsilon > 0$ for algorithm termination.

**Step 2**: For $\mu = 1 \ldots K$, update $\begin{bmatrix} \eta_\mu^{(\tau+1)} \\ \sigma_\mu^{(\tau+1)} \end{bmatrix} = \begin{bmatrix} \eta_\mu^{(\tau)} \\ \sigma_\mu^{(\tau)} \end{bmatrix} - J^{-1}\left(\eta_\mu^{(\tau)}, \sigma_\mu^{(\tau)}\right) \cdot \begin{bmatrix} g_\eta(\eta_\mu^{(\tau)}, \sigma_\mu^{(\tau)}) \\ g_\sigma(\eta_\mu^{(\tau)}, \sigma_\mu^{(\tau)}) \end{bmatrix}$, where the derivation details of $g_\eta\left(\eta_\mu^{(\tau)}, \sigma_\mu^{(\tau)}\right), g_\sigma\left(\eta_\mu^{(\tau)}, \sigma_\mu^{(\tau)}\right)$ and $J^{-1}\left(\eta_\mu^{(\tau)}, \sigma_\mu^{(\tau)}\right)$ are given in Appendix. A.

**Step 3**: If $\left|\eta_\mu^{(\tau+1)} - \eta_\mu^{(\tau)}\right| \geq \varepsilon$ or $\left|\sigma_\mu^{(\tau+1)} - \sigma_\mu^{(\tau)}\right| \geq \varepsilon$, $\mu = 1 \ldots K$, update $\tau = \tau+1$ and go to Step 2; otherwise, terminate the algorithm.

With the model parameters estimated by the above procedure, $\widehat{\mathbf{\Theta}} = \{\hat{\eta}_\mu, \hat{\sigma}_\mu\}_{\mu=1}^K$, we calculated the probability of being discharged to disposition $\mu$, $\widehat{F}_\mu(t \mid \widehat{\mathbf{\Theta}})$ as:

$$\widehat{F}_\mu(t \mid \widehat{\mathbf{\Theta}}) = \Pr(T_{i,\mu} \leq t) = \Phi\left(\frac{\ln t - \hat{\eta}_\mu}{\hat{\sigma}_\mu}\right), \mu = 1 \ldots K. \quad (4)$$

Based on the estimated latent survival model, there is a need to further develop a sampling algorithm to generate predictive samples of NH residents, including how long each resident will stay in NH and which disposition each resident will be discharged to. There is a lack of such sampling algorithm in the existing literatures of healthcare simulation. We developed a sampling algorithm to account for the multiple discharge dispositions. The proposed algorithm can be summarized as follows.

**Step 1**: Calculate disposition-specific discharge risk $\hat{d}_\mu(t \mid \widehat{\mathbf{\Theta}}) = \frac{\phi\left(\frac{(\ln t - \hat{\eta}_\mu)}{\hat{\sigma}_\mu}\right)}{t \hat{\sigma}_\mu \Phi\left(-\frac{(\ln t - \hat{\eta}_\mu)}{\hat{\sigma}_\mu}\right)}, \mu = 1 \ldots K$

**Step 2**: Simulate NH LOS $\widehat{T}_i$ for individual $i$ based on total risk of being discharged $\hat{d}(t \mid \widehat{\mathbf{\Theta}}) = \sum_\mu \hat{d}_\mu(t \mid \widehat{\mathbf{\Theta}})$

**Step 3**: Simulate discharge disposition $Z_{i\mu}$ for resident $i$ based on categorical distribution with probability $p_{Z_{i\mu}=\mu} = \frac{\hat{d}_\mu(\widehat{T}_i \mid \widehat{\mathbf{\Theta}})}{\hat{d}(\widehat{T}_i \mid \widehat{\mathbf{\Theta}})}, \mu = 1 \ldots K$. Individual $i$ stayed in NH till the dwell duration reached $\widehat{T}_i$, and would be discharged to disposition $Z_{i\mu}$.

## 2.2 Individual Daily Staff-Time Simulator

During the NH stay, the individual daily service need may significantly differ from one resident to another, due to the diverse chronic conditions (e.g., vascular disease, osteoporosis, dementia and depression), multifunctional (e.g., physical and cognitive) limitations, and different types of therapies (e.g., audiology, occupational and/or physical therapy) and treatments (e.g., radiation, dialysis and/or skin treatment) received. As a result, the individual service demand measured by the amount of daily staff-time needed, was significantly diverse among NH residents. Such demand heterogeneity was not well addressed in the existing literatures. In many of existing works of simulation-based decision support platform for healthcare systems [11-12, 15, 22, 24], they often characterized service demand of patients based on patient volume without differentiating the service demand difference among different individuals. To account for daily service demand heterogeneity among NH residents, we developed a service need-based simulator via incorporating domain knowledge and existing quantitative studies [25, 27].

The proposed simulator consisted of two submodules: service need classification and staff-time generation. In the submodule of service need classification, we classified a heterogeneous population of NH residents into multiple service need groups based on their varied individual characteristics. The daily service demand of NH residents from different service need groups, quantified as daily staff-time needed (in minutes), was characterized by different distribution parameters. To be more specific, we utilized RUG-IV [27], the most recent version of a patient classification system adopted by the CMS for reimbursement purpose. This classification categorized NH residents into multiple service need groups and each service need group comprised residents with similar resource usage level. The classification was performed on individual health records extracted from the Minimum Data set (MDS). MDS was established as part of a federally mandated process for standardized and comprehensive assessment of all Medicare and Medicaid paid NH residents in CMS certified NHs throughout the U.S. [28]. The raw MDS data contained admission and discharge information of each resident, as well as rich resident-level health assessment information with more than 600 coded items, such as individual demographics, functional performance and disease diagnoses. MDS data was primarily utilized for Medicare and Medicaid reimbursement, personalized care planning, and NH quality monitoring.

To reduce the dimensionality of the classification, we generated 9 composite variables to capture different aspects of NH residents, including their physical, mental and cognitive functioning performance (e.g., $x_1$, $x_2$ and $x_9$), illnesses and comorbidities (e.g., $x_6$, $x_7$, $x_8$), as well as service level on rehabilitation, restorative care and extensive medical service (e.g., $x_3$, $x_4$, $x_5$). Table 1 showed a detailed description of the 9 composite variables. For each resident, we calculated the value of each composite variable from the raw MDS data. For instance, $x_1$ was Activities of Daily Living (ADLs), which measured the physical disability condition of NH residents [29]. ADL score ranged from 0 to 16 and was calculated from resident-level assessment of four "late loss" ADLs, namely, bed mobility, transfer, toilet use, and eating. A higher ADL value implied a higher level of functional assistance required by the resident. When generating each hypothetical resident (i.e., agent) in the simulation, we assigned the above 9 attributes (composite variables) and classified the generated agent into a service need group accordingly. Fig. 1 (a) further elaborated this classification process. Attributes (e.g., $x_1$, $x_2$, etc.) annotated along different branches and a set of rules described in each diamond symbol determined the pathways of NH residents being classified into different service need groups. For example, Fig. 1 (b) gave a more detailed view of the classification rules of the first diamond symbol, which were established by attributes $x_1$ (ADL score), $x_4$ (rehabilitation service level) and $x_5$ (extensive medical service level). NH residents who needed daily living assistance with different levels, rehabilitation with different intensity levels and extensive medical service would be classified into one of the mutually exclusive service need groups (e.g., groups 1 to group 9).

**Table 1**
Summary of variables used in service need classification [29]

| Variable | Type | Range | Description |
|---|---|---|---|
| $x_1$ | Ordinal | 0-16 | Activities of Daily Living (ADL) score: Functional disability |
| $x_2$ | Binary | 0, 1 | Depression indicator: Whether resident has depressive symptoms |
| $x_3$ | Binary | 0, 1 | Restorative nursing services indicator: Whether resident receives restorative assistance of motion, splint/brace, bed mobility, etc. |
| $x_4$ | Ordinal | 0-5 | Rehabilitation services level: level of therapy services needed for speech, audiology, occupational and physical therapy. |
| x$_5$ | Ordinal | 0-3 | Extensive medical services level: level of medical services needed on tracheostomy, ventilator and/or respirator treatment. |
| x$_6$ | Binary | 0, 1 | High special care indicator: Whether resident has septicemia, diabetes, quadriplegia, COPD, fever, etc. |
| x$_7$ | Binary | 0, 1 | Low special care indicator: Whether resident has cerebral palsy, Parkinson's disease, multiple sclerosis, oxygen therapy, etc. |
| x$_8$ | Binary | 0, 1 | Clinical complex indicator: Whether resident has pneumonia, hemiplegia, surgical wounds, burns, chemotherapy, etc. |
| x$_9$ | Binary | 0, 1 | Cognitive impairment indicator: Whether resident has mental score <=9, hallucinations, delusions, behavioral symptoms, etc. |

With the above submodule of service need classification, we further used daily staff-time needed to quantify service demand of NH residents in each service need group, as demonstrated in the subsequent submodule of staff-time generation. Due to the lack of actual staff-time measurements in our studied NH, we incorporated STRIVE project from existing NH studies to quantify the "representative" daily staff-time needed (in minutes) of each type of caregivers (e.g., nurses or aides) for NH residents in each service need group. STRIVE, one of the most recent national staff-time projects, provided nationwide reference values of average daily staff-time needed for residents in each service need group based on the aggregation of raw staff-time measurements (collected using personal digital assistants) of approximately 97,000 NH residents from more than 200 representative high-quality NHs across different states. For each service need group, STRIVE contained (i) daily average staff-time spent directly with or on behalf of a resident; and (ii) daily average staff-time proportion spent indirectly for supporting the delivery of care for a resident. The former was defined as direct care staff-time, which involved activities such as feeding, helping dress, giving medications, charting for a resident, calling a physician about a resident, etc. The latter was defined as indirect care staff-time, which involved activities such as stocking medication cabinet, performing administration, participating in training sessions, taking time for breaks and meals, etc. To generate daily staff-time needed for each NH resident, we denoted $T_{kig}^{\text{direct}}$ and $T_{kig}^{\text{indirect}}$ as staff-time of direct care and indirect care of type $k$ caregivers for resident $i$ in service need group $g$, respectively. To ensure the modeling simplicity and practical convenience [30], we considered exponential distribution with single parameter to model daily staff-time of direct care and indirect care, and further assumed they were independent. Specifically, we assumed $T_{kig}^{\text{direct}} \sim \text{Expo}(\lambda_{k1g})$ and $T_{kig}^{\text{indirect}} \sim \text{Expo}(\lambda_{k2g})$, where $1/\lambda_{k1g}$ and $1/\lambda_{k2g}$ were nationwide reference values of daily average staff-time of type $k$ caregivers for NH residents in service need group $g$ extracted from STRIVE project. The total daily staff-time of resident $i$ then became $T_{kig} = T_{kig}^{\text{direct}} + T_{kig}^{\text{indirect}}$, which was simulated on a daily basis from hypo-exponential distribution, i.e., $T_{kig} \sim \text{HypoExpo}(\lambda_{k1g}, \lambda_{k2g})$. The individual daily staff-time on each type of caregiver was repeatedly simulated for each resident until the end of his/her NH stay (see Fig. 1).

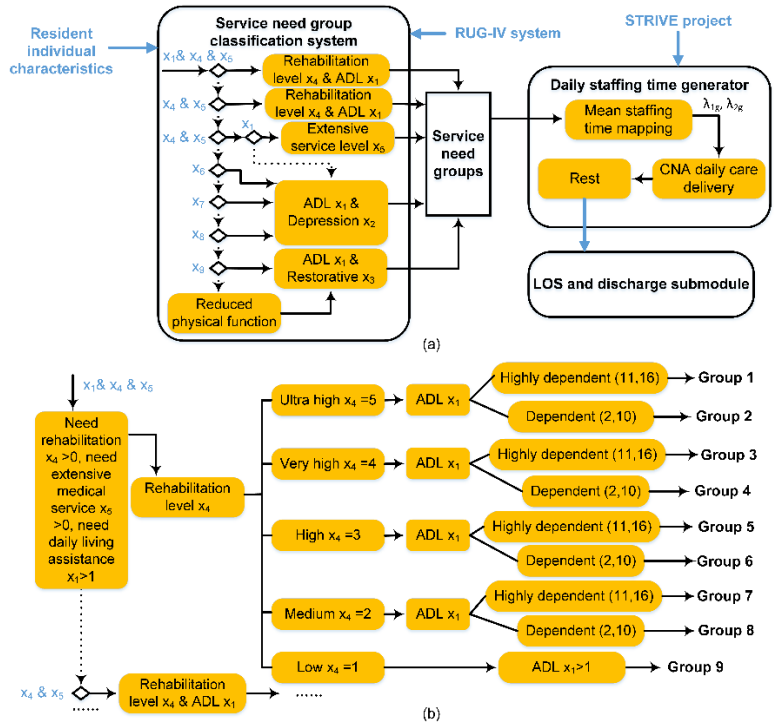

Fig. 1. Individual daily staff-time simulator, (a) overview, (b) zoom-in details of classification rules of first diamond symbol

### 2.3 Simulation Implementation and Graphical User Interface Development

We developed the above modules and implemented the proposed simulation platform in AnyLogic simulation environment. Fig. 2 illustrated a descriptive diagram of the proposed predictive analytics integrated computer simulation platform. We first assigned a set of attributes to each NH individual to represent his/her physical/mental status and service category (e.g., rehabilitation, restoration, etc.). Then, we simulated resident arrival, dwell and discharge based on the individual attributes. These attributes also jointly determined the course of resident's states over time and the daily staff-time needed on each type of caregivers. The aggregated daily service demand at facility level over time must be met by the staffing supply from nursing staffs assigned. The resultant labor cost was computed to assess the decision performance of each staffing strategy at facility level. Note that a promising staffing strategy needed to be robust to resident census fluctuation at facility level and staffing need randomization at individual level.

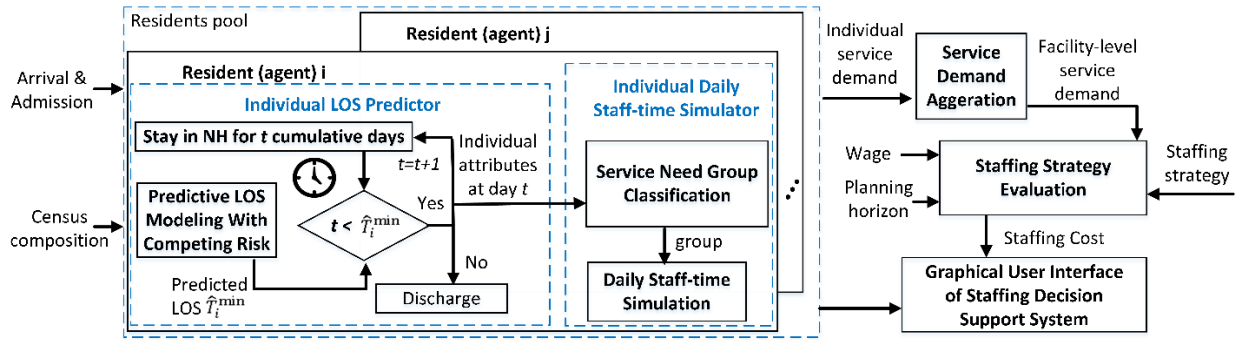

Fig. 2. Descriptive diagram of the developed simulation platform

To improve the usability of our tool in real practice by NH administrators, we further developed a user-friendly graphical user interface (GUI), as shown in Fig. 3. It included three panels: (1) input setting panel,

(2) visualization panel, and (3) evaluation output panel. The input setting panel allowed the user to specify simulation settings, such as number of simulation days, arrival rate of residents, parameters settings of the LOS model, and SR ratio to be evaluated. It also allowed for importing individual characteristics data of NH residents. The visualization panel gave a descriptive statistical summary of simulated residents in terms of their characteristics and LOS quantities. The evaluation output panel displayed the facility-level service demand over time under some scenario of resident census, the staffing level over time, and the total labor cost of the strategy evaluated.

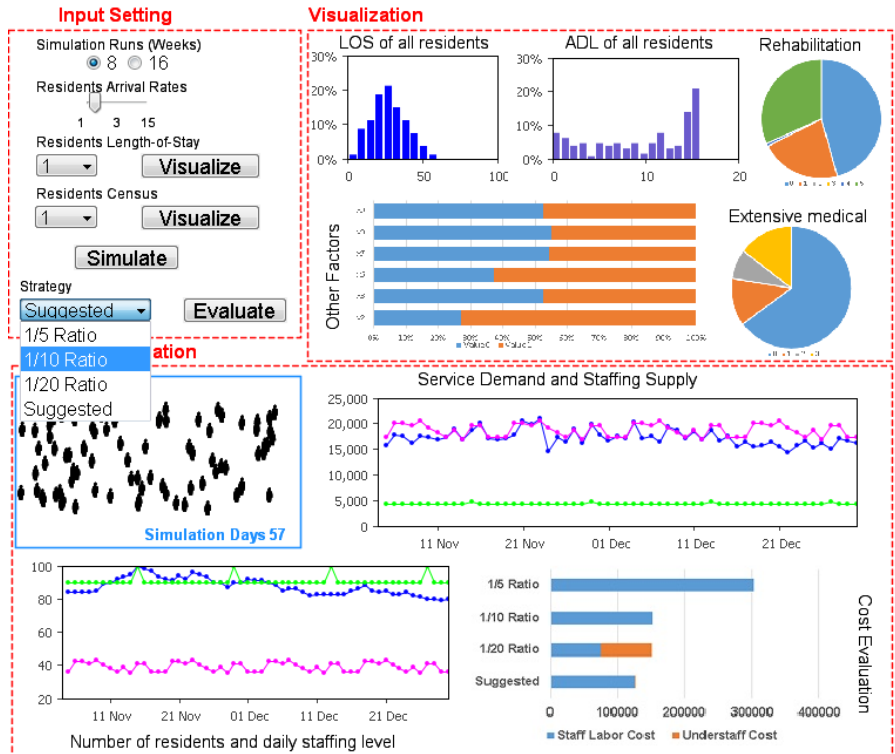

Fig. 3. GUI of the decision support platform

## 3. Real Case Study

### 3.1 Real Data Description

To demonstrate the viability of the simulation in real-world settings, we obtained Minimum Data Set (MDS) data from a representative NH facility of our industrial collaborator in the Tampa Bay area. A summary of descriptive statistics was shown in Table 2. More than half of the residents were females and the majority were elderly residents over 65 years old. Most of the residents (over 90%) had at least one chronic disease (e.g., cancer, hypertension and diabetes). Typically, 75% of NH residents required daily living assistance at a medium level (e.g., ADL from 2 to 10) while 20% of residents were more physically disabled and required daily living assistance at a higher level (e.g., ADL from 11 to 16). More than 95% residents received rehabilitation and therapy service at various intensity levels.

**Table 2**
Summary of individual characteristics in real NH data

| Characteristics | Type | Statistics |
|---|---|---|
| Age | Numeric | Mean: 76, SD: 10.73 |
| Gender, n (%) | Binary | |
|     Female | | 64.2% |
| Chronic diseases, n (%) | Binary | |
|     Having at least one disease | | 91.6% |
| ADL, n (%) | Ordinal | |
|   Less dependent (<2) | | 5% |
|   Medium (2-10) | | 75% |
|   More dependent(11-16) | | 20% |
| Rehabilitation level, n (%) | Ordinal | |
|   No need of rehabilitation (0) | | 5% |
|   Low/Medium rehabilitation (1-2) | | 18% |
|   High rehabilitation (3-5) | | 77% |
| Extensive medical care level, n (%) | Ordinal | |
|   No need of extensive care (0) | | 96% |
|   Need any of the following: tracheostomy care, ventilator or respirator care, isolation for active infectious disease (1-3) | | 4% |
| Having cognitive impairment, n (%) | Binary | 12.7% |

### 3.2 Prediction Performance Comparison and Model Validation

Based on the NH data acquired from 759 NH residents, we analyzed daily number of arrivals based on parametric distribution using real arrival data. We compared various distributions such as Poisson distribution Pois($\lambda$) and Negative binomial distribution, i.e., NB($r$, $p$). We employed Chi-square test to evaluate the goodness-of-fit of the fitting distributions. The arrival model based on NB($r$, $p$) distribution exhibited the best goodness-of-fit performance. The p-value of the estimated NB($\hat{r}, \hat{p}$) was 0.38, which indicated satisfactory goodness-of-fit of real arrival data. For details of estimated parameters (including point estimates and standard errors), please refer to Appendix B.

    With the estimated arrival model, we then employed the proposed latent survival model to analyze real NH LOS data and evaluated its prediction performance. In the real data, community and hospital were the two main discharge dispositions (about 61% of residents went back to community and 24% of residents were readmitted to hospital). Other discharge dispositions, such as transferring to another NH or death, were negligible and thus discarded. We evaluated the prediction performance of the proposed model. As illustrated in Fig. 4 (a), the predicted Kaplan-Meier (K-M) curve of the predictive samples of the proposed approach was close to the K-M curve of the observed actual LOS samples. We further compared the prediction performance of the proposed model with two alternative modeling approaches [9, 26], namely, LOS modeling without considering multiple discharge dispositions, and LOS modeling by segregating LOS data into multiple groups based on the observed labels. As shown in Figs. 4 (b) and (c), both alternative approaches exhibited unsatisfactory prediction performance and they both tended to underestimate LOS as compared to the proposed model. The inaccurate prediction of LOS model without considering multiple discharge dispositions was attributed to its simplified modeling assumption of utilizing identical model parameters in capturing NH stay with essentially different types of dispositions. The modeling approach of segregating LOS data into multiple groups based on the observed disposition labels and estimating model via group-specific data also yielded less prediction accuracy since it neglected the latent discharge events. A resident segregated into community discharge group will still have potential risk of being transiting to hospital and vice versus. Thus, partitioning LOS observations into multiple groups cannot reflect multiple types of discharge dispositions and latent discharge events associated with NH stay of each resident.

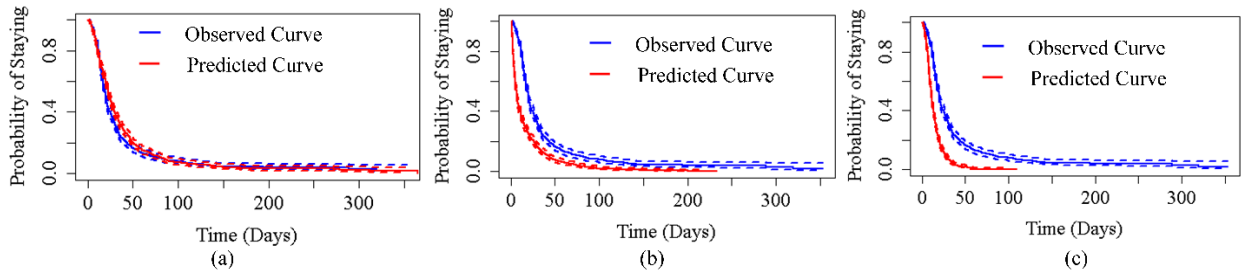

Fig. 4. Comparison between the observed (blue) and the predicted (red) survival curves based on different modeling approaches, (a) proposed model, (b) LOS model without considering multiple discharge dispositions, (c) LOS model by segregating data into multiple groups based on the observed labels

With the demonstrated superior performance of proposed predictive LOS model, we further validated the performance of simulation outputs. We compared the simulated samples of daily resident volume over a month with the actual observed samples, as shown in Fig. 5 (a). It visually verified that the simulated results exhibited a similar distribution to the real observations. In addition, two-sample Kolmogorov-Smirnov (K-S) test was also used to compare their differences and a p-value of 0.52 implied that there was no statistically significant difference between the two. On the other side, the simulation output of conventional approach without considering multiple discharge dispositions did not show satisfactory validation results (see Fig. 5 (b)) with a p-value of < 2.2e-16 in the K-S test. The simulated resident volume based on the conventional LOS models tended to be lower than the actual resident volume. This was attributed to the underestimated LOS of conventional models (as shown in Fig. 4) which neglected the multiple discharge dispositions. The LOS underestimation resulted in a higher-than-actual NH discharge rate and a lower-than-actual resident volume. To further provide validation of the simulation outputs at finer scale, we compared the simulated daily number of residents over time (under multiple replication runs) based on the proposed approach with real data. As shown in Fig. 5 (c), the 95% confidence bands of simulated resident volume can fully cover the observed daily number of residents over time.

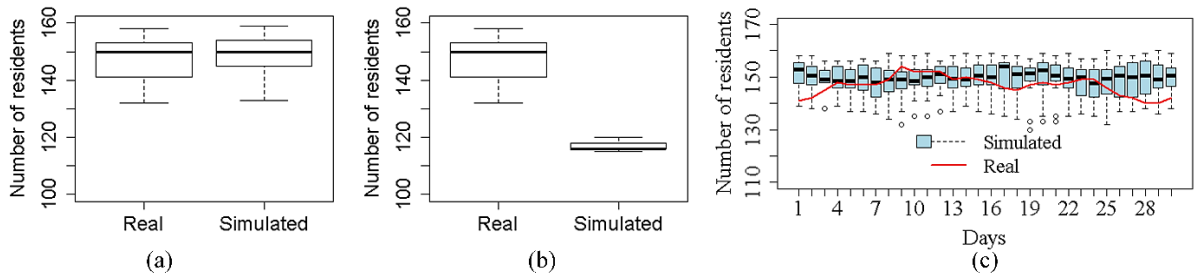

Fig. 5. Comparison between observed data and simulated samples of daily number of residents (a) based on the proposed LOS model, (b) based on LOS model without considering multiple discharge dispositions, (c) over time at finer scale based on the proposed LOS model

To further emphasize the importance of the proposed predictive analytics method, we compared the simulated demand, staffing supply and total labor cost based on LOS model with and without considering multiple discharge dispositions, as shown in Fig. 6. As compared to the proposed LOS model with improved prediction performance, the prediction inaccuracy of LOS model which neglected multiple discharge dispositions propagated to facility-level and induced the lower-than-actual service demand generated in Fig. 6 (a), and subsequently resulted in the inadequate staff supply determined in Fig. 6(b). The improper staffing decision based on LOS model without considering multiple discharge dispositions eventually led to a higher labor cost due to significant understaffing cost incurred, as shown in Fig. 6(c). In summary, with the improved modeling accuracy of proposed predictive analytics integrated simulation model, the suggested staffing strategy could meet the residents demand at reduced overall labor cost.

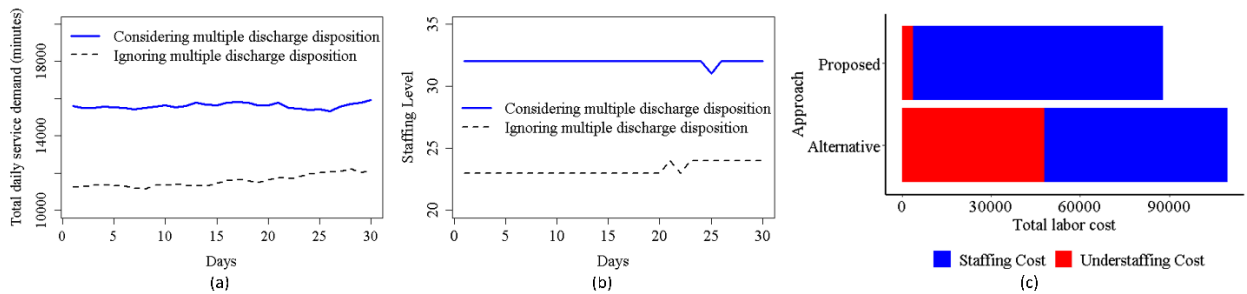

Fig. 6. Facility-level performance comparison of (a) simulated service demand, (b) staffing supply, and (c) labor cost between LOS models with (proposed) and without (alternative) considering multiple discharge dispositions

### 3.3 Staffing Strategy Evaluation and Comparison

After validating the simulation model, we can then evaluate various staffing strategies with different staff-to-resident (SR) ratios, i.e., strategies that maintain a pre-specified SR ratio. We primarily compared three staffing strategies, namely, (1) the minimum SR ratio based on state regulation (i.e., state SR ratio), (2) the SR ratio implemented in current NH based on observed staffing data (i.e., facility SR ratio), and (3) the promising SR ratio recommended by the proposed platform (i.e., suggested SR ratio). We considered the total labor cost during a simulation period as performance metric in evaluating and comparing different staffing strategies. Given the simulated facility-level service demand over time and a staffing strategy with specified SR ratio, we can calculate the total staffing surplus in minutes (in the case that overstaffing occurred) or the total unmet demand in minutes (in the case that understaffing occurred). To satisfy the unmet demand, additional temporary nursing staffs (e.g., PRN nurses as needed or agency aides) needed to be called in. The total labor cost over a simulation period embraced both the total planned staffing cost in paying wages for regular nursing staffs over the whole simulation period, and the total understaffing staffing cost in paying wages for temporary nursing staffs over the whole simulation period.

To simulate the service demand for evaluating different SR ratios, we considered a real NH resident census based on the acquired data (see Table 2). We first considered certified nursing assistants (CNAs), or nursing aides, since they were essential nursing staff who provided the most direct care to residents. Other types of caregivers will be considered in the following subsections. With the above performance evaluation scheme, we compared and illustrated three staffing strategies of CNAs, namely, (1) state SR ratio, (2) facility SR ratio, and (3) suggested SR ratio identified by the developed simulation platform. As shown in Fig. 7, the suggested SR ratio provided a more appropriate level of staff resource commitment in meeting the service demand, as compared to the other two staffing strategies. Note that the state SR ratio led to understaffing whereas the facility SR ratio led to overstaffing.

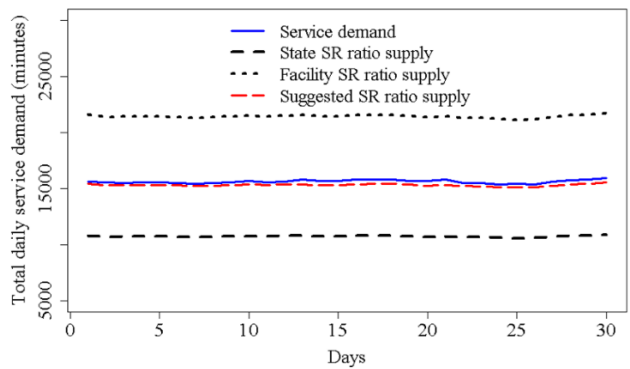

Fig. 7. Matching CNA service demand and supply under different SR ratios

Fig. 8 illustrated the total labor cost of CNAs under different SR ratios, including the above three. The total labor cost included planned CNA staffing cost and additional understaffing cost due to CNA shortage. Due to either understaffing or overstaffing, both state and facility SR ratios yielded a larger total labor cost.

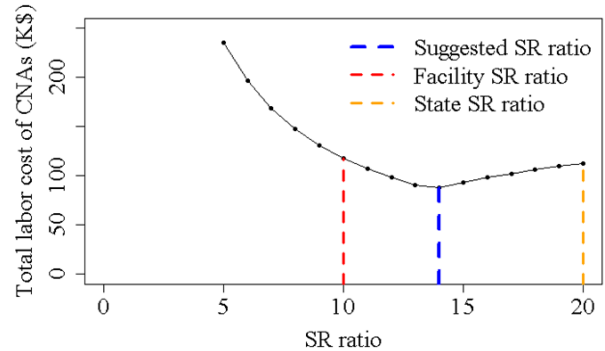

Fig. 8. Comparison of total labor cost of CNAs under different SR ratios

To account for uncertainty in our stochastic simulation, we ran 20 replications, which was determined by the approximation formula in [31]. We further increased number of replications (e.g., 30 and 50) to empirically investigate the variability of simulation outputs as well as computational time. The variability of simulation results under different replications scenarios were similar. However, when more replications were considered, the computational time dramatically increased. Thus, 20 replications were finally selected. Based on the simulated service demand, we then evaluated the staffing strategies and summarized the results in Table 3, including both point and interval estimates of the total labor cost. The discrepancy between service demand and supply, quantified by average daily staff-time associated with overstaffing and understaffing, was also reported. As shown in Table 3, the suggested SR ratio achieved a lower total labor cost of CNAs with reduced overstaffing and understaffing.

**Table 3**
Total labor cost of CNAs under different staffing ratios with 20 replications

| Staffing strategy (SR ratio) | Total labor cost (thousand $) | 95% CI (thousand $) | Staffing cost (thousand $) | Avg. daily overstaffing (minutes) | Avg. daily understaffing (minutes) |
|---|---|---|---|---|---|
| State SR ratio (1/20) | 110.2 | (107.3, 112.9) | 58.1 | 0 | 2366.6 |
| Facility SR ratio (1/10) | 116.2 | (114.4, 117.9) | 116.2 | 2914.4 | 0 |
| Suggested SR ratio (1/14) | 85.9 | (83.8, 88.1) | 82.9 | 32 | 135.3 |

### 3.4 What-if Scenario with Different Censuses

The NH resident census may vary considerably at different time periods (e.g., season, year) or across different NHs, which may lead to adjustment of SR ratios. The developed simulation was able to generate "what-if" scenarios of alternative censuses and to evaluate the SR ratios under each scenario. In the above scenario of real census (baseline scenario), 75% and 95% of NH residents were at a medium level of physical dependency (e.g., ADL score between 1 and 12) and needed therapy services, such as physical, occupation and/or speech therapy, respectively. We considered three alternative scenarios as follows.

- "More physically independent" scenario (S1): 70% of residents were less physically dependent (i.e., ADL between 0 and 1) and could either live independently or require less daily living assistance. The mean ADL decreased by 60% compared to the baseline scenario.
- "Less physically independent" scenario (S2): 70% of residents were more physically dependent (i.e., ADL between 11 and 16) and required more daily living assistance. The mean ADL increased by 60% compared to the baseline scenario.
- "Less therapy needs" scenario (S3): Residents had same ADL to those in the baseline scenario. However, the percentage of residents who received therapy service decreased by 50%.

The simulated service demand under each scenario and staffing supply with different SR ratios are illustrated in Fig. 9. It is noticed that, under different scenarios, resident volume was assumed approximately identical (i.e., based on the same arrival distribution and LOS model). Conventional volume-based service demand modeling failed to consider individual difference and consequently cannot capture the service demand difference among different scenarios with different census compositions. In the proposed work, we characterized heterogeneous service demand of NH residents based on the varied individual characteristics. We were able to generate service demand under various census compositions of a heterogeneous population of NH residents. Fig. 9 showed that service demand among different scenarios differed significantly. Under these different scenarios, NH residents had various health conditions and service needs, and it required adaptive adjustment of staffing strategy to meet the diverse service demand. The one-size-fits-all staffing strategy based on a fixed SR ratio could not uniformly meet the NH service demand under different scenarios. For instance, the state SR ratio (i.e., 1 CNA for 20 residents) could more adequately meet the service demand under scenario S1 than the facility SR ratio (i.e., 1/10). It is because most residents under scenario S1 were more independent and consequently fewer CNAs were needed. On the contrary, under scenario S2, when most residents were more physically dependent, the amount of staffing based on the state SR ratio became inadequate and was less appropriate than the facility SR ratio to address the higher service demand from more frail and dependent NH residents.

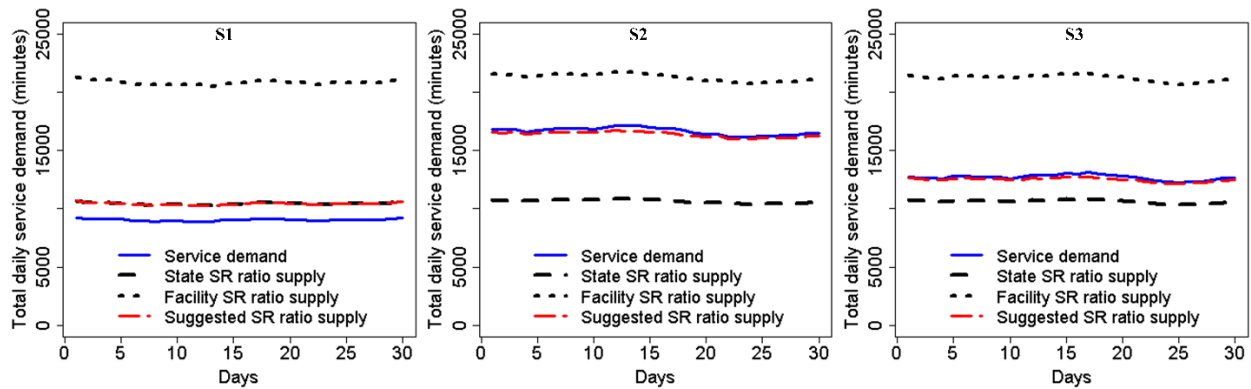

Fig. 9. Comparison of service demand under different composition scenarios and staffing supply based on staffing strategies with different SR ratios

To meet the service demand under different census scenarios, we identified the suggested SR ratios and compared their corresponding total labor costs of CNAs under the same scale. As shown in Fig. 10, the suggested SR ratio under S1 was lower than that under the baseline scenario. It is because residents under S1 were more physically independent, and thus requiring fewer CNAs to assist with daily living of activities. On the other hand, the suggested SR ratio under S2 was higher because residents were more physically disabled. As compared to the baseline scenario, the suggested SR ratio under S3 was lower because even though residents had the same ADL, fewer of them needed rehabilitation service. The reduced rehabilitation service need led to reduced CNA staff-time and thus reduced the number of CNAs needed. Note that in a typical NH, CNAs are not only responsible for providing daily living assistance for residents, they also assist residents with their rehabilitation plans, such as assisting residents with physical or occupational therapy activities established by the therapists.

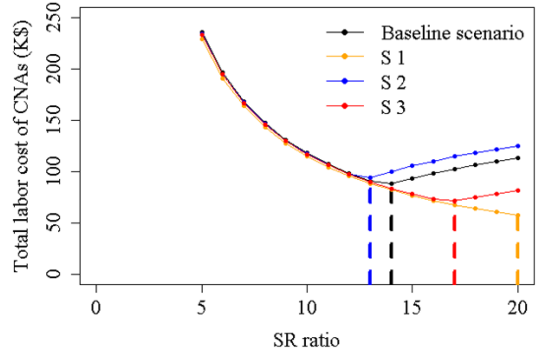

Fig. 10. Suggested SR ratios under different census composition scenarios

Further, we ran 20 replications and calculated the average total labor cost over the replications based on the identified ratios under each census scenario, as summarized in Table 4. The suggested SR ratios under different census scenarios differed due to the service demand variation under different scenarios. More specifically, given the same number of residents, the total amount of service demand and the number of CNAs needed would vary due to different health conditions under different scenarios. The simulation characterized the service demand heterogeneity at a higher fidelity to make more appropriate staffing recommendation at reduced labor cost. It was also noticed that, under all scenarios, the suggested SR ratio yielded a smaller total labor cost, while a one-size-fits-all staffing strategy, such as the one using the state or facility SR ratio, led to a higher labor cost. The higher cost is attributed to inappropriate staffing for either increased understaffing or unnecessarily planned staffing.

**Table 4**
Comparison of total labor cost of CNAs under different composition scenarios

| No. | Staffing strategy (SR ratio) | Total labor cost (thousand $) | 95% CI (thousand $) | Staffing cost (thousand $) | Avg. daily overstaffing (minutes) | Avg. daily understaffing (minutes) |
|---|---|---|---|---|---|---|
| S1 | State SR ratio (1/20) | 58.4 | (57.6, 59.3) | 58.4 | 682.3 | 0 |
|  | Facility SR ratio (1/10) | 116.9 | (115.2, 118.5) | 116.9 | 5994.7 | 0 |
|  | Suggested SR ratio (1/20) | 58.4 | (57.6, 59.3) | 58.4 | 682.3 | 0 |
| S2 | State SR ratio (1/20) | 127.7 | (125.6, 129.7) | 59.4 | 0 | 3101.7 |
|  | Facility SR ratio (1/10) | 118.9 | (117.7, 119.9) | 118.9 | 2300.7 | 0 |
|  | Suggested SR ratio (1/13) | 95.9 | (94.3, 97.5) | 91.4 | 12.3 | 205 |
| S3 | State SR ratio (1/20) | 82.7 | (80.8, 84.6) | 59.1 | 0 | 1072 |
|  | Facility SR ratio (1/10) | 118.2 | (116.6, 119.8) | 118.2 | 4300.4 | 0 |
|  | Suggested SR ratio (1/17) | 72.5 | (70.9, 74.1) | 69.5 | 11.7 | 135.6 |

### 3.5 Staffing Evaluation of Other Types of NH Caregivers

To further demonstrate the flexibility of the proposed work, we extended the simulation platform to staffing evaluation of other types of caregivers in NH, such as registered nurse (RN) and licensed practical nurse (LPN). Unlike CNAs who are responsible for caring NH residents via providing personal assistance and support on daily routine activities, both RNs and LPNs have more advanced training and provide more advanced nursing service than CNAs. RNs are primarily responsible for administering medication and treatment to residents, coordinating care plans, operating medical equipment and educating residents as well as their families. Under supervision of RNs or physicians, LPNs mainly provide basic and routine medical care for residents, such as recording vital signs, monitoring health conditions of residents and administering injections. Due to the different roles of each type of caregivers, the corresponding service demand (quantified in daily staff-time need) will also differ. As shown in Fig. 11 (a) and (b), we evaluated and compared different staffing strategies for RN and LPN. The staffing strategies in comparison included state SR ratio, facility SR ratio and suggested SR ratio identified by the proposed platform. As shown in Fig. 11, both state and facility SR ratios for both RN and LPN led to significant understaffing.

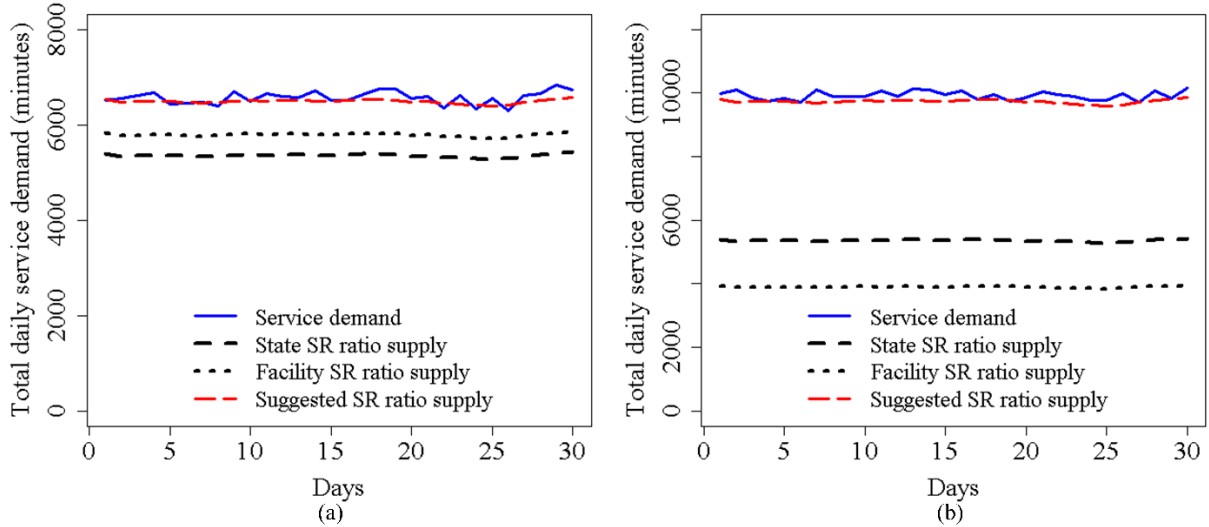

Fig. 11. Matching service demand and supply under different SR ratios for (a) RNs, and (b) LPNs

Fig. 12 (a) and (b) further showed the reduced total labor cost of the suggested SR ratio as compared to the cost of the other SR ratios for RN and LPN, respectively. Further, Table 5 summarized both the point and interval estimates of total labor cost under different SR ratios for RN and LPN. Due to the understaffing, both state and facility SR ratios led to significantly larger total labor cost for both RN and LPN.

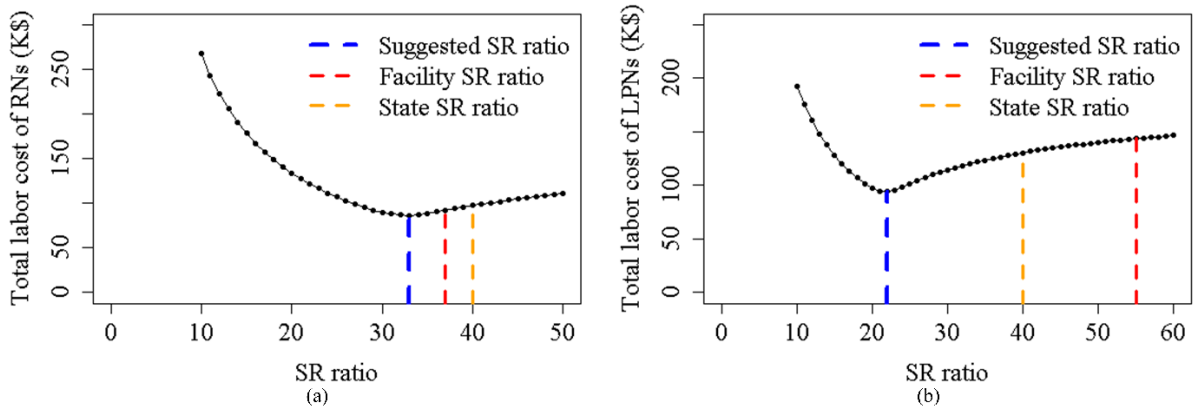

Fig. 12. Comparison of total labor cost of (a) RNs, and (b) LPNs, under different SR ratios

**Table 5**
Total labor cost of RNs and LPNs under different staffing ratios

| Care-giver type | Staffing strategy (SR ratio) | Total labor cost (thousand $) | 95% CI (thousand $) | Staffing cost (thousand $) | Avg. daily overstaffing (minutes) | Avg. daily understaffing (minutes) |
|---|---|---|---|---|---|---|
| RN | State (1/40) | 97.4 | (94.7, 100.1) | 66.9 | 0 | 608.9 |
|  | Facility (1/37) | 92.1 | (89.5, 94.6) | 72.4 | 2.2 | 394.1 |
|  | Suggested (1/33) | 86.6 | (84.7, 88.5) | 81.1 | 68.4 | 109.4 |
| LPN | State (1/40) | 130.5 | (120.8, 132.7) | 48.2 | 0 | 2286.8 |
| | Facility (1/55) | 143.7 | (141.2, 146.1) | 35.1 | 0 | 3017 |
| | Suggested (1/22) | 94.1 | (92.6, 95.6) | 87.6 | 83.8 | 180.1 |

## 4. CONCLUSION AND FUTURE WORK

In this paper, a data-driven simulation tool was proposed to characterize heterogeneous service demand of NH residents and to evaluate system performance outcomes (e.g., facility-level labor cost of CNAs) under different SR ratios. In the proposed simulation, a predictive LOS model considering multiple discharge dispositions was first developed based on latent survival analysis to estimate dwelling duration of NH residents with improved prediction performance. Further, a predictive analytics method integrated computer simulation model was developed to characterize heterogeneous service demand of individual NH resident based on the varied individual characteristics and to generate the daily staff-time needed via incorporating domain knowledge from a national study. Next, the daily staff-time was aggregated at the facility level. Our study based on a realistic NH resident population helped justify the validity of our tool for the following analysis requests. First, given a sample of residents with various individual characteristics, the tool can output service demand over time and evaluate the corresponding labor costs at the facility level under a specific SR ratio. Second, based on the simulated service demand, the tool can further evaluate different SR ratios and thus identify a suggested ratio with the most reduced labor cost. Third, the tool allows users to perform "what-if" analysis by simulating service demand under different scenarios of NH resident composition and identifying the most cost-saving SR ratio under each scenario.

In the proposed simulation, model inputs, such as LOS quantity (modeled by latent survival analysis) and arrival process (modeled with a negative binomial distribution), were modeled based on the resident data of a single NH available to us. Thus, our simulation could not be directly applied to different NHs with different facility and resident characteristics without re-estimating the inputs. However, the input estimation procedure (i.e. parameter estimation and distribution selection) could still be viable. For instance, the formulation and estimation of the predictive LOS model based on latent survival analysis in Equations 1 and 2 are generic and can be applied to any finite number of discharge dispositions, not necessarily restricted to two dispositions as in this paper. Different parametric models other than log-normal distribution may apply as long as they exhibit reasonable goodness-of-fit for the new dataset. In addition, our simulation platform utilized the national staffing time measurement study (i.e., STRIVE) to characterize daily service demand of NH residents in each service need group. Ideally, such staff-time generation submodule, including the staff-time distribution parameters, needed to be recalibrated with detailed staff-time measurements from local NHs to improve the modeling accuracy. However, due to the lack of local staff-time measurement study available, there may exist modeling bias with the use of data from a national study. We would actively collaborate with local NHs in future to address this data challenge. Moreover, in this study, we investigated the aggregated staffing decision and assessed the facility-level performance with current available data. We would collect more data to investigate more detailed staffing decisions such as shift scheduling and to explore the individual-level decision in capacity management (e.g., individual difference in required number of beds and number of staffs). Besides, we will study observed interactions among residents and interactions between caregivers and residents within same NH facility, and analyze system nonlinearity due to unobserved interaction between individual and organization.

# APPENDICES

## A. Derivation details of $g_\eta\left(\eta_\mu^{(\tau)}, \sigma_\mu^{(\tau)}\right)$, $g_\sigma\left(\eta_\mu^{(\tau)}, \sigma_\mu^{(\tau)}\right)$ and $J(\eta_\mu^{(\tau)}, \sigma_\mu^{(\tau)})$

$g_\eta\left(\eta_\mu^{(\tau)}, \sigma_\mu^{(\tau)}\right)$ can be explicitly calculated as

$$g_\eta\left(\eta_\mu^{(\tau)}, \sigma_\mu^{(\tau)}\right) = \frac{1}{\sigma_\mu^{2(\tau)}} \sum_{i \in I_\mu} \left(\ln t_i - \eta_\mu^{(\tau)}\right) + \sum_{i \notin \cup_\mu I_\mu} \frac{q_i\left(\eta_\mu^{(\tau)}, \sigma_\mu^{(\tau)}\right)}{\sigma_\mu^{(\tau)}}$$

$g_\sigma\left(\eta_\mu^{(\tau)}, \sigma_\mu^{(\tau)}\right)$ can be explicitly calculated as

$$g_\sigma\left(\eta_\mu^{(\tau)}, \sigma_\mu^{(\tau)}\right) = \sum_{i \in I_\mu} \left(-\frac{1}{\sigma_\mu^{(\tau)}} + \frac{1}{\sigma_\mu^{3(\tau)}}\left(\ln t_i - \eta_\mu^{(\tau)}\right)^2\right) + \sum_{i \notin \cup_\mu I_\mu} \frac{q_i\left(\eta_\mu^{(\tau)}, \sigma_\mu^{(\tau)}\right)\left(\ln t_i - \eta_\mu^{(\tau)}\right)}{\sigma_\mu^{2(\tau)}}$$

where $q_i\left(\eta_\mu^{(\tau)}, \sigma_\mu^{(\tau)}\right) = \dfrac{\phi\left(\frac{\ln t_i - \eta_\mu^{(\tau)}}{\sigma_\mu^{(\tau)}}\right)}{\Phi\left(-\frac{\ln t_i - \eta_\mu^{(\tau)}}{\sigma_\mu^{(\tau)}}\right)}$

$J(\eta_\mu^{(\tau)}, \sigma_\mu^{(\tau)})$ can be expressed as $J(\eta_\mu^{(\tau)}, \sigma_\mu^{(\tau)}) = \begin{bmatrix} A_{11} & A_{12} \\ A_{12} & A_{22} \end{bmatrix}$

where

$$A_{11} = -\sum_{i \in I_\mu}\left(\frac{1}{\sigma_\mu^{2(\tau)}}\right) + \frac{1}{\sigma_\mu^{2(\tau)}} \sum_{i \notin \cup_\mu I_\mu} q_i(\eta_\mu^{(\tau)}, \sigma_\mu^{(\tau)})\left(\frac{\ln t_i - \eta_\mu^{(\tau)}}{\sigma_\mu^{(\tau)}} - q_i(\eta_\mu^{(\tau)}, \sigma_\mu^{(\tau)})\right)$$

$$A_{12} = -\frac{2}{\sigma_\mu^{3(\tau)}}\sum_{i \in I_\mu}\left(\ln t_i - \eta_\mu^{(\tau)}\right) + \frac{1}{\sigma_\mu^{2(\tau)}}\sum_{i \notin \cup_\mu I_\mu} q_i(\eta_\mu^{(\tau)}, \sigma_\mu^{(\tau)})$$

$$\cdot \left(\left(\frac{\ln t_i - \eta_\mu^{(\tau)}}{\sigma_\mu^{(\tau)}}\right)^2 - 1 - \frac{(\ln t_i - \eta_\mu^{(\tau)}) \cdot q_i(\eta_\mu^{(\tau)}, \sigma_\mu^{(\tau)})}{\sigma_\mu^{(\tau)}}\right)$$

$$A_{22} = \sum_{i \in I_\mu}\left(\frac{1}{\sigma_\mu^{2(\tau)}} - \frac{3(\ln t_i - \eta_\mu^{(\tau)})^2}{\sigma_\mu^{4(\tau)}}\right) + \frac{1}{\sigma_\mu^{2(\tau)}}\sum_{i \notin \cup_\mu I_\mu} \frac{(\ln t_i - \eta_\mu^{(\tau)}) \cdot q_i(\eta_\mu^{(\tau)}, \sigma_\mu^{(\tau)})}{\sigma_\mu^{(\tau)}}$$

$$\cdot \left(\left(\frac{\ln t_i - \eta_\mu^{(\tau)}}{\sigma_\mu^{(\tau)}}\right)^2 - 2 - \frac{(\ln t_i - \eta_\mu^{(\tau)}) \cdot q_i(\eta_\mu^{(\tau)}, \sigma_\mu^{(\tau)})}{\sigma_\mu^{(\tau)}}\right)$$

## B. Detailed NH inputs and practical knowledge used for the simulation platform and case study

Table B.1  Estimated model parameters based on real NH data

| Model | Parameter | Description | Estimate | Standard Error |
|---|---|---|---|---|
| Arrival | $r$ | size of arrival event | 4.95 | 1.1144 |
|  | P | probability of arrival | 0.64 | 0.0524 |
| LOS | $\eta_1$ | community-specific mean on log scale | 3.41 | 0.0377 |
|  | $\sigma_1$ | community-specific standard deviation on log scale | 0.94 | 0.0299 |
|  | $\eta_2$ | hospital-specific mean on log scale | 4.52 | 0.1072 |
|  | $\sigma_2$ | hospital-specific standard deviation on log scale | 1.58 | 0.0835 |